\documentclass{article}
\usepackage{spconf,amsmath,graphicx,multirow}

\usepackage{url}

\title{Few-shot Learning for CT Scan based COVID-19 Diagnosis}
\twoauthors
 {Yifan Jiang \qquad Han Chen \qquad Hanseok Ko\sthanks{This material is based upon work supported by the Air Force Office of Scientific Research under award number FA2386-19-1-4001. (Corresponding author: Hanseok Ko. E-mail: hsko@korea.ac.kr)}}
	{School of Electrical and Computer Engineering\\
	 Korea University\\
	 Seoul, Korea}
 {David K. Han}
	{Department of Electrical and \\ Computer Engineering\\
	Drexel University\\
	Philadelphia, PA 19104, USA}

\begin{document}
%\ninept
%
\maketitle

\begin{abstract}
Coronavirus disease 2019 (COVID-19) is a Public Health Emergency of International Concern infecting more than 40 million people across 188 countries and territories. Chest computed tomography (CT) imaging technique benefits from its high diagnostic accuracy and robustness, it has become an indispensable way for COVID-19 mass testing. Recently, deep learning approaches have become an effective tool for automatic screening of medical images, and it is also being considered for COVID-19 diagnosis. However, the high infection risk involved with COVID-19 leads to relative sparseness of collected labeled data limiting the performance of such methodologies. Moreover, accurately labeling  CT images require expertise of radiologists making the process expensive and time-consuming. In order to tackle the above issues, we propose a supervised domain adaption based COVID-19 CT diagnostic method which can perform effectively when only a small samples of labeled CT scans are available. To compensate for the sparseness of labeled data, the proposed method utilizes a large amount of synthetic COVID-19 CT images and adjusts the networks from the source domain (synthetic data) to the target domain (real data) with a cross-domain training mechanism. Experimental results show that the proposed method achieves state-of-the-art performance on few-shot COVID-19 CT imaging based diagnostic tasks.
\end{abstract}

\section{Introduction}
\label{sec:intro}

Coronavirus disease 2019 (COVID-19) \cite{covid19} is an ongoing global pandemic that was declared by the World Health Organization (WHO) on 11 March 2020. It has already infected more than 40 million individuals and caused 1,119,369 death, as of 20 October 2020 \cite{covid19_num}. COVID-19 is highly contagious and it spreads more readily compared to similar infectious diseases such as Middle East Respiratory Syndrome (MERS) or Severe Acute Respiratory Syndrome (SARS) \cite{mahase2020coronavirus}. To slow down the rapid transmission of this disease, it is necessary to detect the COVID-19 in an early stage of infection.

With the emergence of deep learning, medical imaging area also benefited from effective feature representation capability of deep learning techniques \cite{kamnitsas2017efficient,yang2019deep,fan2020inf,li2020artificial,kang2020diagnosis}. However, applying deep learning on COVID-19 diagnosis is challenging due to lack of sufficiently large labeled data, particularly of COVID-19 CT data, as it involves high infection risk and the labeling process requiring experienced radiologists \cite{ai2020correlation}. To enable effective deep learning based COVID-19 diagnosis, it is necessary to develop a novel approach capable of learning in few-shot conditions (only limited data is available).

One promising approach that can tackle the above issue is to use synthetic data for model training. However, a model trained with synthetic data may not perform satisfactorily on real data when applied directly. This is because of the domain shift problem: synthetic data (source domain) may not necessarily have similar distribution compared to the distribution of real data (target domain). To handle the domain shift problem, some supervised domain adaptation (SDA) methods are proposed recently. FADA \cite{motiian2017few} applied adversarial learning to learn embedding features that maximize the distance between two domains while aligning on a semantic level. CCSA \cite{motiian2017unified} proposed a series of loss functions in order to manage the domain gap for a few-shot domain adaptation tasks. \textit{d}-SNE \cite{xu2019d} introduced a new approach that exploits the stochastic neighborhood embedding theory and modified-Hausdorff distance to improve the few-shot classification performance. Although, many efforts have been done on SDA or few-shot COVID-19 diagnosis areas \cite{chen2020momentum,ter2020single}, applying domain adaptation on CT images for the COVID-19 diagnostic task is relatively a new area, and our proposed method is one of the first attempts in utilizing synthetic chest CT scans for few-shot COVID-19 diagnostic task. 

In this paper, we propose a novel supervised domain adaptation based few-shot COVID-19 diagnostic method applied to CT scans. The proposed method consists of a Siamese structure, and the domain shift problem is solved with a cross-domain training mechanism. The main idea is to learn a model that can quantify three distances at domain-level: (a) Classification loss $\mathcal{L}_{c}$ to maximize the distribution distance between samples from different categories; (b) Cross-domain pairing loss $\mathcal{L}_{cp}$ to minimize the distribution distance between samples from a different domain but of the same category; (c) Cross-domain detaching loss $\mathcal{L}_{cd}$ to maximize the distribution distance between samples from different domains and categories. 

\begin{figure}[!ht]
\centering
\includegraphics[width=8cm]{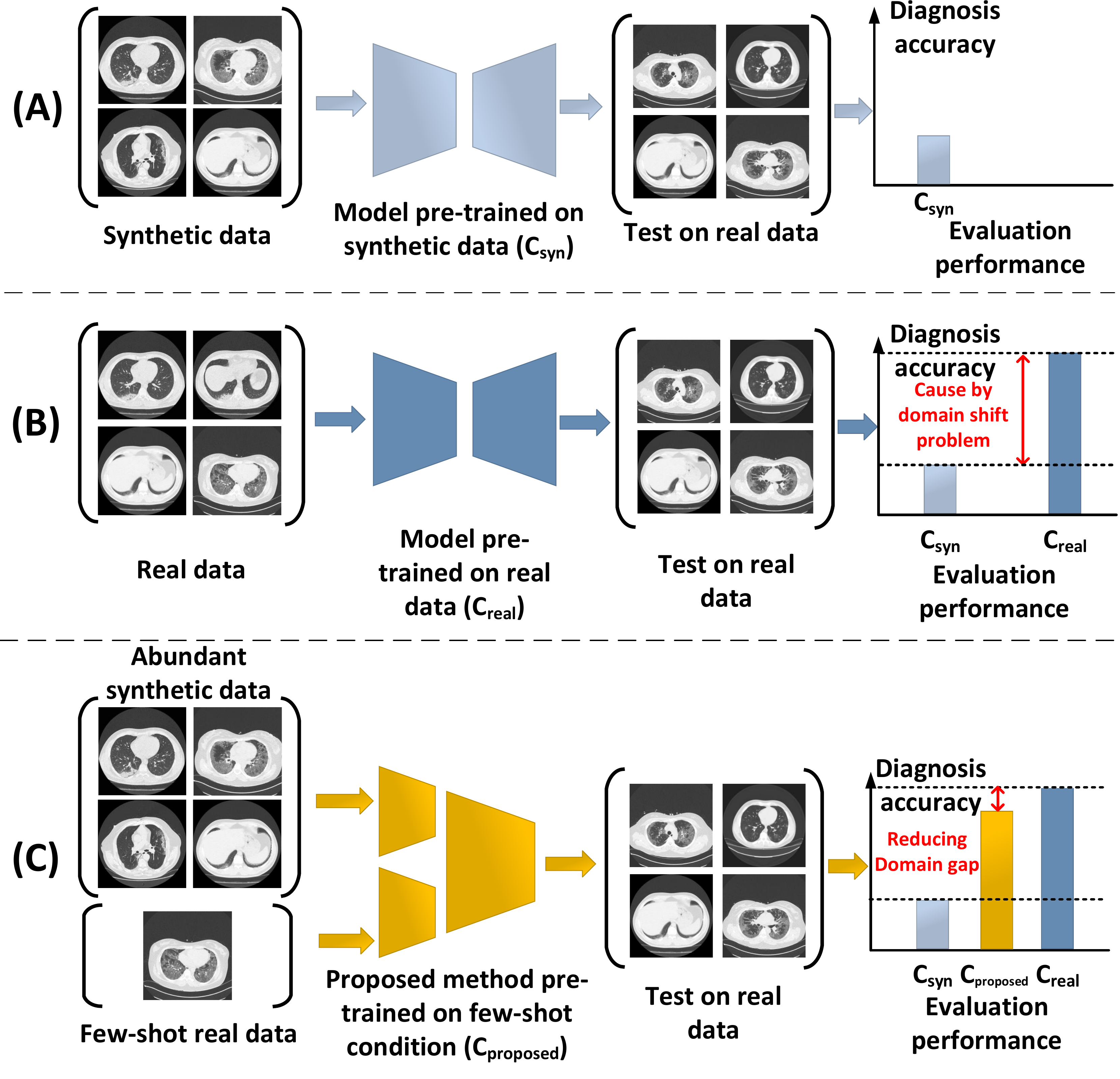}
\caption{Demonstration of domain shift problem and proposed solution. (A) Model trained on a large synthetic dataset then tested on real data; (B) Model trained on a large real dataset then tested on real data; (C) The proposed model trained on a large synthetic dataset and few real data, then evaluated on real data.}
\label{fig1} % this gives the figure a unique name that you can refer to in the main text \ref{fig:...}
\end{figure}

An illustration of the domain shift problem and the proposed solution is shown in Figure \ref{fig1}. Figure \ref{fig1} (A) depicts the situation in which a COVID-19 CT based diagnostic model is pre-trained by a large amount of synthetic data and tested on real data. The model in Figure \ref{fig1} (B), on the other hand, was given a large quantity of real data for training and was tested the same way as in the model in Figure \ref{fig1} (A). As would be expected, Figure \ref{fig1} (A) model performs poorly due to the domain shift problem compared to Figure \ref{fig1} (B) model. Figure \ref{fig1} (C) shows the proposed method which utilizes the same synthetic data as in model (A) plus few real data to reduce the domain gap between synthetic data and real data so that it can achieve a similar performance level as in (B), as synthetic data can easily be generated from our previous work \cite{jiang2020covid}. The main contributions of our work are as follows:

\begin{itemize}

\item [(1)] We propose a novel chest CT data based COVID-19 diagnostic method designed for few-shot conditions in which only a small quantity of COVID-19 CT data is available. To the best of our knowledge, the proposed method stands the first domain adaptation method that utilizes synthetic COVID-19 CT data for a few-shot COVID-19 diagnostic task. 
\item [(2)] We propose a Siamese network structure that is trained by a novel cross-domain training mechanism. This cross-domain training mechanism enables an effective domain transfer via three different losses ($\mathcal{L}_{c}$, $\mathcal{L}_{cp}$ and $\mathcal{L}_{cd}$) in few-shot condition.
%\item [(3)] Since proposed method shows a promising performance on few-shot COVID-19 CT diagnosis task, it provides some potentials in practical applications, for instance, the mass test of COVID-19 or other novel diseases during very early stage when only few data is available. 

\end{itemize}

\section{Proposed method}
\label{sec:proposed}

\begin{figure}[!ht]
\centering
\includegraphics[width=8cm]{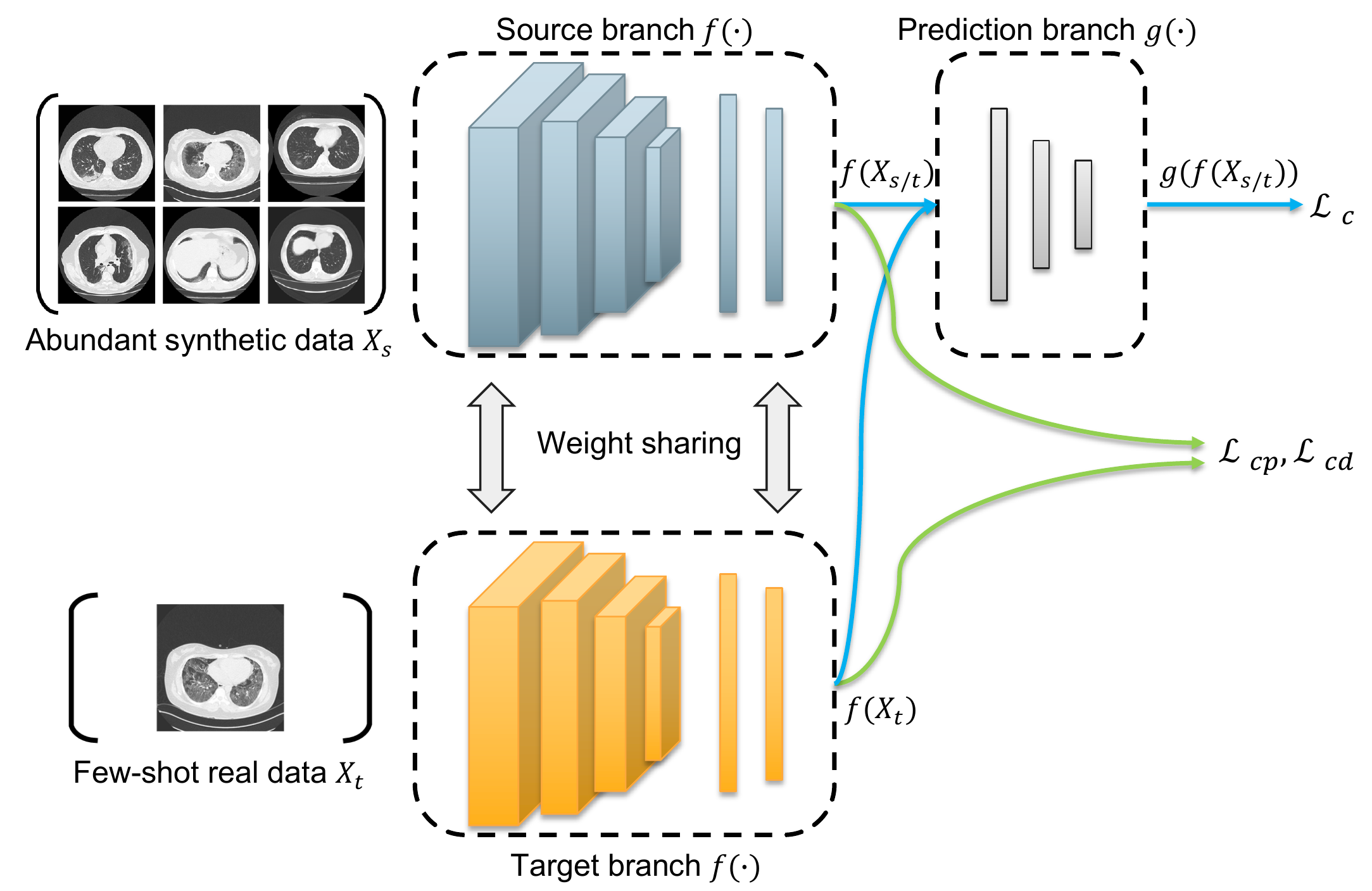}
\caption{Overview of the proposed method. The proposed model mainly consists of three parts: the source branch $f(\cdot)$ depicted in blue color, the target branch $f(\cdot)$ in orange color, and the prediction branch $g(\cdot)$ in gray color. The cross-domain losses ($\mathcal{L}_{cc}$ and $\mathcal{L}_{ci}$) and the classification loss $\mathcal{L}_{c}$  are derived through the green arrow and the blue arrow, respectively.}
\label{fig2} % this gives the figure a unique name that you can refer to in the main text \ref{fig:...}
\end{figure}

In this work, we propose a novel Siamese network based model for a few-shot COVID-19 CT diagnostic task as illustrated in Figure \ref{fig2}. The Siamese network structure is basically formed in three components: source branch $f(\cdot)$, target branch $f(\cdot)$ and prediction branch $g(\cdot)$.  Source and target branches have the same network structure which consists of a feature extractor and two fully-connected (FC) layers. The prediction branch is a network that contains three FC layers. During the training stage, weight sharing occurs between source and target branches as they take staggered input of synthetic and real as $(X_{s1}, X_{t1}, X_{s2}, X_{t2}, ......, X_{sN}, X_{tN},)$. Since the synthetic data outnumber the real data in a large proportion, the real data were reused. Two embedded feature vectors ($f_s(X_s)$ and $f_t(X_t)$) are extracted through the two branches. Only $f_s(X_s)$ is passed to the prediction branch for calculating classification loss $\mathcal{L}_{c}$ while both $f_s(X_s)$ and $f_t(X_t)$ are used to compute the cross-domain losses ($\mathcal{L}_{cp}$ and $\mathcal{L}_{cd}$). The classification loss and the cross-domain loss are used together to construct the overall loss for updating the network. During the test stage, a real CT image is passed through the network, and the network makes a binary diagnostic decision.

\subsection{Classification loss}                                             In order to train the proposed classifier to classify an input CT scan to be positive or negative, we propose a classification loss $\mathcal{L}_{c}$ as follows:
\begin{equation}
\begin{aligned}
\mathcal{L}_{c}=-[y\cdot \log g(f(x))+(1-y)\log(1-g(f(x)))]
\end{aligned}
\label{eq1}
\end{equation}
where $x$ denotes the input CT scan, $y$ denotes the binary label $(0, 1)$ of the corresponding input. The binary category cross entropy loss learns the difference between positive case and negative case and teaches the network to recognize the characteristics of the lesion representation associated with COVID-19. Since the data distributions of source and target domains are different, this domain gap can influence the diagnostic performance of the model when the network is pre-trained in the source domain but is tested in the target domain. Therefore, the classification loss alone is not sufficient to handle the domain shift problem, and further cross-domain measures are required to deal with the domain shift problem.

\subsection{Cross-domain pairing loss}

We define here a novel Cross-domain pairing loss for managing the distance between features from different domains but have the same label. The cross-domain pairing loss is defined as

%Cross-domain losses are proposed for improving the cross-domain performance on few-shot COVID-19 CT diagnosis task. To be specific, cross-domain losses handle the situation that training a diagnosis model on efficient synthetic data and limited real data, then testing on real data. In the case of cross-domain paring loss, this loss manages the distance between the features which come from different domains but have the same label. 

\begin{equation}
\begin{aligned}
\mathcal{L}_{cp}=D(p(f(X_{s}^{p}), p(f(X_{t}^{p}))))+D(p(f(X_{s}^{n}), p(f(X_{t}^{n}))))
\end{aligned}
\label{eq2}
\end{equation}
where $X_{s}^{p}$ represents a sample from the source domain with a positive label, while $X_{t}^{n}$ denotes a sample from the target domain with a negative label. $D$ is the distance between two probability distributions, $p(\cdot)$, and it is computed by average pairwise Euclidean distances between points of the same label from the two domains. By applying the cross-domain pairing loss $\mathcal{L}_{cp}$, the model can learn the pair-wise (same category) relationship between two domains by minimizing the distance between the two feature distributions. 

\subsection{Cross-domain detaching loss}
In order to further enhance the cross-domain diagnostic performance, we propose a cross-domain detaching loss, aimed at maximizing the distance between two feature distributions of different classes. The definition of cross-domain detaching loss is defined as follows:

\begin{equation}
\begin{aligned}
\mathcal{L}_{cd}=D(p(f(X_{s}^{p}), p(f(X_{t}^{n}))))+D(p(f(X_{s}^{n}), p(f(X_{t}^{p}))))
\end{aligned}
\label{eq3}
\end{equation}
Similar to $\mathcal{L}_{cp}$, the cross-domain detaching  $\mathcal{L}_{cd}$ uses Euclidean distance to manage the difference between the two feature distributions. The learning object is to maximize $\mathcal{L}_{cd}$ so that the diagnostic model is able to effectively separate the distributions well at the feature-level for enhanced performance.

\subsection{Overall learning objective}
The overall learning object is defined as

\begin{equation}
\begin{aligned}
\mathcal{L}_{overall}=\mathcal{L}_{c}+\alpha(\mathcal{L}_{cp}-\mathcal{L}_{cd})
\end{aligned}
\label{eq4}
\end{equation}
where hypo-parameter $\alpha$ denotes a weight factor of the cross-domain losses. By applying both the classification loss $\mathcal{L}_{c}$ and the cross-domain losses $\mathcal{L}_{cp} , \mathcal{L}_{cd}$, our proposed COVID-19 diagnostic model can not only effectively classify the positive/negative cases within the domain, but also can transfer the knowledge from the source domain to the target domain. Thus, the proposed combination of the loss functions fully exploits the large number of synthetic data for COVID-19 CT diagnostic task when only a small number of real data are given.

\section{Experiments}
\label{sec:exp}

\subsection{Experimental settings}

\noindent \textbf{Dataset.} We constructed our dataset by using the data from both the source and the target domains. The source domain data is generated by our previous work \cite{jiang2020covid}, and the target domain data comes from a public COVID-19 CT dataset which contains 29 individual cases \cite{covid19_data}. Here, all the CT slices are divided into training set (20) and test set (9) by patient level. Specifically, we apply a combination of 6,000 source domain slices (synthetic data) and 60 target domain slices (real data) to form the training set, and we use 600 real CT scans as our test set. The COVID-19 diagnostic task is formulated as a binary classification task here, therefore, there are only two possible categories: positive and negative. In order to evaluate the proposed model, we randomly select \textit{n} positive cases and \textit{n} negative cases from the target domain slices and pair them with a randomly selected source domain group which contains 600 samples, so we can obtain $2n\cdot 600$ source-target pairs for the n-shot learning task.

\noindent \textbf{Evaluation metrics.} We report the diagnostic performance by two metrics: accuracy and F1 score. We randomly re-sample 10 times for building 10 individual training sets, and report the results with the format as MEAN$\pm$95\% CONFIDENCE INTERVAL among the 10 folds.

% Accuracy is the most common used metric for image classification task,
% it simply measures the percentage of the cases which are correctly classified. However, the accuracy metric is not enough for evaluating a machine leering on medical imaging area. We consider F1 score as a metric in this paper to evaluate the sensitivity and robustness of proposed method. F1 score can balance between precision and recall metrics, so it can be considered as suitable metric for COVID-19 CT diagnosis task. To enhance the rigor and persuasiveness of experimental results, 

% \begin{equation}
% \begin{aligned}
% F1\:score=2\cdot\frac{Precision\cdot Recall}{Precision+Recall}
% \end{aligned}
% \label{eq5}
% \end{equation}

\noindent \textbf{Experimental details.} All CT scans are transformed to gray images on a Hounsfield unit (HU) scale [-600,1500] and resized to $512\times512$. The learning rate is set as 0.001 with a decay rate 0.95. The weight factor $\alpha$ is 0.25.

% We finish all the experiments in an Ubuntu 18.04 environment using an Intel i7 9700k CPU and a GeForce RTX Titan GPU with Keras 2.4.3, using TensorFlow 2.2.0 as a backend.  

\begin{table}[]
	\centering
	\caption{Diagnostic performance comparison of few-shot COVID-19 CT diagnostic task (n-shot: 5-shot, feature extractor: Xception \cite{chollet2017xception}, the best evaluation score is marked in bold. Higher number of the metrics is better.)}
	\begin{tabular}{|c|c|c|}
	    \hline
            Methods & Metrics & Performance\\ 
		\hline
            \multirow{2}{*}{OURS} & Accuracy & \textbf{0.8040$\pm$0.0356} \\ 
        \cline{2-3}
            & F1 score & \textbf{0.7998$\pm$0.0384}\\ 
		\hline
		    \multirow{2}{*}{Source only} & Accuracy & 0.5353$\pm$0.0268 \\ 
        \cline{2-3}
            & F1 score & 0.5133$\pm$0.0272\\ 
    	\hline
    	    \multirow{2}{*}{CCSA \cite{motiian2017unified}} & Accuracy & 0.7068$\pm$0.0443 \\ 
    	\cline{2-3}
            & F1 score & 0.6841$\pm$0.0573\\ 
    	\hline
    	    \multirow{2}{*}{FADA \cite{motiian2017few}} & Accuracy & 0.7300$\pm$0.0688 \\ 
    	\cline{2-3}
            & F1 score & 0.7167$\pm$0.0796\\ 
    	\hline
    	    \multirow{2}{*}{\textit{d}-SNE \cite{xu2019d}} & Accuracy & 0.7693$\pm$0.0415 \\ 
    	\cline{2-3}
            & F1 score & 0.7651$\pm$0.0426\\     
    	\hline
	\end{tabular}
\label{table1}
\end{table}

\subsection{Experimental results}
In this sub-section, we focus on comparing the performance between the proposed method and other state-of-the-art supervised domain adaptation methods, including CCSA \cite{motiian2017unified}, FADA \cite{motiian2017few}, \textit{d}-SNE \cite{xu2019d}. In order to show how does our domain adaptation help to improve the cross-domain diagnostic performance, we also involve a \textit{source only} competitor which is trained on a deep network with only source domain data and tested on a target domain test set.  This experiment is a 5-shot learning task and we used the Xception \cite{chollet2017xception} network as the feature extractor. As shown in Table \ref{table1}, the proposed method outperforms the other state-of-the-art supervised domain adaptation approaches on both accuracy and F1 score metrics.  

%The proposed method has much better diagnosis performance than the source only competitor and the other methods as the cross-domain training mechanism reduces the domain gap between source and target domain effectively, and helps to boost the diagnosis performance.

\subsection{Ablation Study of the Proposed Cross-domain Losses}
We discuss the ablation study focused on examining the effectiveness of the components of our proposed loss functions in this sub-section.

\begin{table}[]
	\centering
	\caption{Ablation study for cross-domain losses (n-shot: 5-shot, feature extractor: Xception \cite{chollet2017xception}, the best evaluation score is marked in bold. Higher number of the metrics is better.)}
	\begin{tabular}{|c|c|c|}
	    \hline
            Loss terms & Metrics & Performance\\ 
		\hline
            \multirow{2}{*}{$\mathcal{L}_{c}+\mathcal{L}_{cp}+\mathcal{L}_{cd}$ (OURS)} & Accuracy & \textbf{0.8040$\pm$0.0356} \\ 
        \cline{2-3}
            & F1 score & \textbf{0.7998$\pm$0.0384}\\ 
% 		\hline
% 		    \multirow{2}{*}{only $\mathcal{L}_{c}$ (Source only)} & Accuracy & 0.5353$\pm$0.0268 \\ 
%         \cline{2-3}
%             & F1 score & 0.5133$\pm$0.0272\\ 
    	\hline
    	    \multirow{2}{*}{$\mathcal{L}_{c}+\mathcal{L}_{cp}$} & Accuracy & 0.6553$\pm$0.0313 \\ 
    	\cline{2-3}
            & F1 score & 0.6487$\pm$0.0622\\ 
    	\hline
    	    \multirow{2}{*}{$\mathcal{L}_{c}+\mathcal{L}_{cd}$} & Accuracy & 0.7142$\pm$0.0650 \\ 
    	\cline{2-3}
            & F1 score & 0.6998$\pm$0.0514\\ 
    	\hline
	\end{tabular}
\label{table2}
\end{table}

Our overall loss consists of three terms: classification loss $\mathcal{L}_{c}$, cross-domain pairing loss $\mathcal{L}_{cp}$ and cross-domain $\mathcal{L}_{cd}$. In order to explore the performance contribution of each loss term, we evaluated the proposed model under four conditions:  $\mathcal{L}_{c}+\mathcal{L}_{cp}$, $\mathcal{L}_{c}+\mathcal{L}_{cd}$ and $\mathcal{L}_{c}+\mathcal{L}_{cp}+\mathcal{L}_{cd}$ (ours). Experiment results are summarized in Table \ref{table2}. By comparing the contribution of each cross-domain loss term, it is clear that both the cross-domain paring and cross-domain detaching losses can help to overcome the domain gap and improve cross-domain diagnostic performance. 

\subsection{Ablation Study of n-shot learning task}

\begin{figure}[!ht]
\centering
\includegraphics[width=7cm]{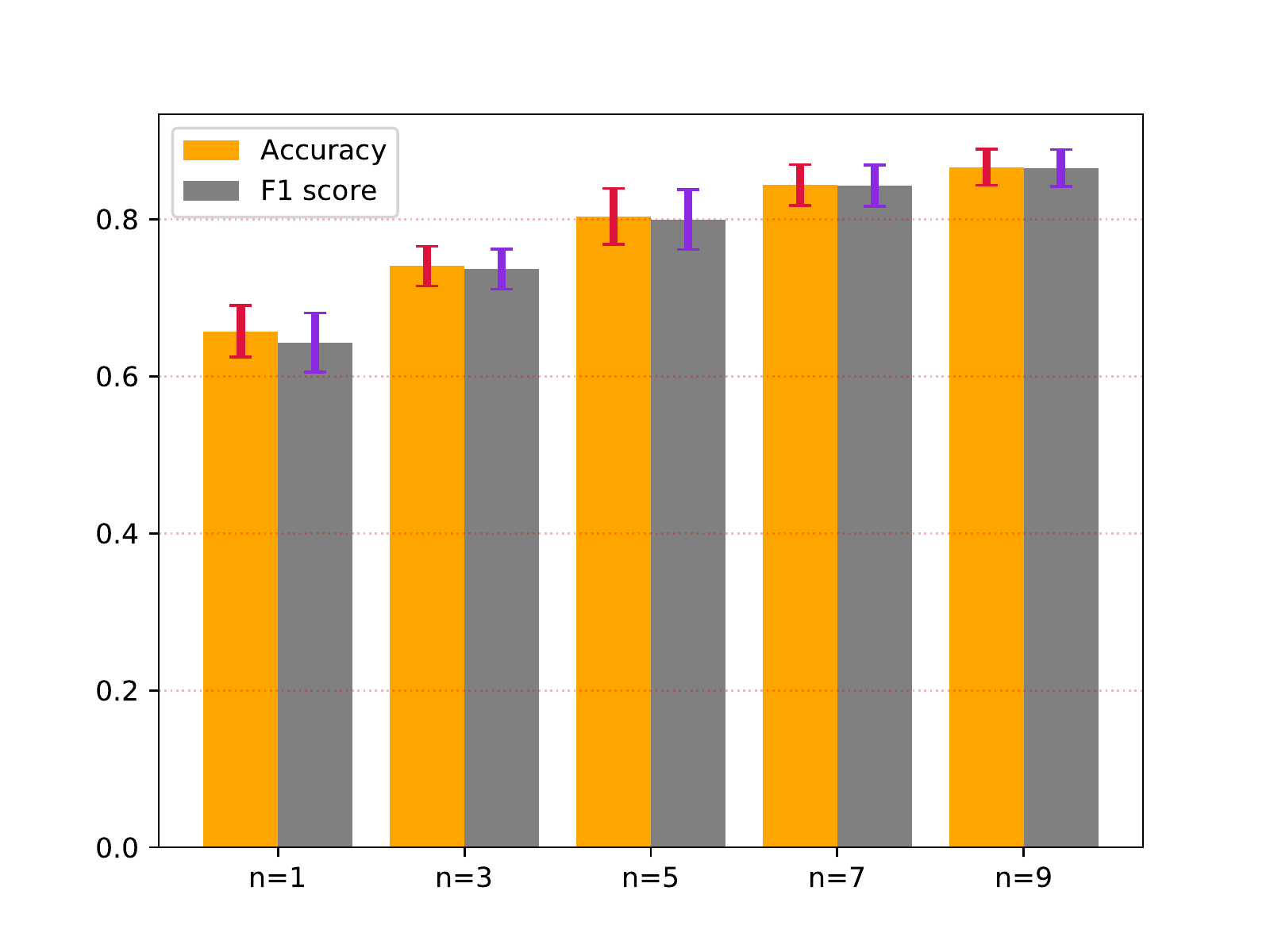}
\caption{Ablation study for n-shot learning task.}
\label{fig3} % this gives the figure a unique name that you can refer to in the main text \ref{fig:...}
\end{figure}

We evaluate the proposed method in terms of its capability in a few-shot learning problem. We consider a total of five cases with \textit{n}=1, 3, 5, 7 and 9, where \textit{n} represents the shot number. Experiment results are shown in Figure \ref{fig3}. From Figure \ref{fig3}, it is clear that the proposed method can effectively handle few-shot diagnoses under diverse n-shot conditions. As expected, the diagnostic performance improves as the shot number $n$ increases. It should be noted, however, that even in the extreme case of \textit{n} = 1, the performance was maintained at above 0.6.

\section{Conclusion and future study}
\label{sec:con}

In this paper, we proposed a supervised domain adaptation based few-shot COVID-19 diagnostic method for CT scans. The novelty of the proposed method consists of constructing a cross-domain training architecture by integrating a Siamese network and introducing two cross-domain training losses in addition to a classification loss. Siamese network based architecture and the proposed cross-domain losses have been demonstrated to be effective in handling the domain shift problem between the source and the target domains. Experimental results on the public COVID-19 CT dataset show that the proposed method outperforms the other state-of-the-art supervised domain adaptation methods on a few-shot COVID-19 CT diagnostic task. For the future plan, we would like to pay attention to channel attention mechanism based COVID-19 diagnostic method using 3D CT volume.

\bibliographystyle{IEEEbib}
\bibliography{strings,refs}

\end{document}